\documentclass[conference]{IEEEtran}
\IEEEoverridecommandlockouts

\usepackage{cite,amsmath,amssymb,amsfonts,algorithm,algpseudocode}
\usepackage{graphicx,epstopdf}
\usepackage{booktabs,tabularx,url}
\usepackage[caption=false,font=footnotesize]{subfig}
\usepackage[bookmarks=false]{hyperref}

\usepackage{multirow}

\usepackage{threeparttable}
\usepackage[inline]{enumitem}

\usepackage{flushend}
\usepackage{balance}

\usepackage{tikz}
\usepackage{pgfplots}
\pgfplotsset{compat=1.18} 

\makeatletter
\newcommand{\linebreakand}{\end{@IEEEauthorhalign}\hfill\mbox{}\par\mbox{}\hfill\begin{@IEEEauthorhalign}}
\makeatother

\begin{document}

\title{
LEO-Aware DRL Meta-Scheduler for \\5G Non-Terrestrial Network Slicing%





}

\author{
    \IEEEauthorblockN{
        Víctor Vilchez\IEEEauthorrefmark{1},
		Tiago P. C. de Andrade\IEEEauthorrefmark{2},
        Edward Hinojosa\IEEEauthorrefmark{1},
        Edmundo Madeira\IEEEauthorrefmark{2},
        and Carlos A. Astudillo\IEEEauthorrefmark{2}
    } 
    \IEEEauthorblockA{
        \IEEEauthorrefmark{1}Professional School of Computing Science, National University of San Agustin, Peru\\
        \IEEEauthorrefmark{2}Institute of Computing, University of Campinas, Brazil\\
    }
    Email: \{vvilchezd, ehinojosa\}@unsa.edu.pe, edmundo@ic.unicamp.br, \{tandrade, castudillo\}@unicamp.br
}

\maketitle
\thispagestyle{empty}


\begin{abstract}
The integration of Low Earth Orbit (LEO) Non-Terrestrial Networks (NTNs) into 5G and upcoming 6G architectures introduces various challenges, including severe propagation delays, ultra-high base station mobility, and channel non-stationarity, complicating radio resource management of heterogeneous network slices. In this paper, we propose a deep reinforcement learning (DRL) meta-scheduler for twin-timescale resource allocation.
Our solution adopts a decoupled Open Radio Access Network (RAN) architecture, in which a strategic 100~ms meta-scheduler selects scheduling policies for the different network slices using stale telemetry, while a fast-timescale MAC packet scheduler processes per-TTI user requests. The resulting Markov Decision Process captures non-stationary orbital dynamics and heterogeneous SLAs 
constraints via a TD3 agent. 
Simulation results under varying traffic load show that, unlike other solutions, the proposed meta-scheduler 
explicitly trades a statistically insignificant 1\% capacity fraction ($p > 0.05$) to strictly bound the variance and overall magnitude of RLC-layer queuing delay for Mission-Critical (MC) traffic. Crucially, it enforces this isolation without inducing the broadband slice starvation characteristic of standard maximum-CQI heuristics, establishing a robust foundation for 6G O-RAN NTN resource allocation.
\end{abstract}

\section{Introduction}
\label{sec:intro}

Non-Terrestrial Networks (NTNs) were introduced in 5G to extend connectivity beyond terrestrial infrastructure, enabling wide-area coverage via satellites and supporting diverse use cases such as broadband access, mission-critical communications, and massive IoT. The reduced latency and high capacity of Low Earth Orbit (LEO) systems make them central to future 6G services, though they also introduce challenges due to ultra-high satellite mobility and rapidly varying channel conditions~\cite{3gpp_2017_studynrntn, 3gpp_2018_solutions}.

Context-aware resource management in LEO NTNs is a key challenge for 6G, as satellite beams must serve heterogeneous slices under fast-changing orbital geometry and traffic demands~\cite{azari_2022_evolution,cao_2024_collaborative}. As satellites move from sparse regions to urban hotspots, path loss can fluctuate by over $20$ dB, Doppler shifts reach kHz, and coverage topology changes within seconds~\cite{liu_2025_reconfigurable}. Meanwhile, the RAN must enforce Service Level Agreements (SLAs)~\cite{ojaghi_2022_slicedran} for eMBB, mMTC, and NTN-adapted delay-sensitive services. Since LEO propagation delays preclude terrestrial sub-millisecond bounds, URLLC SLA objectives shift toward extreme transmission reliability and strictly bounded MAC-layer queuing delay, while respecting 3GPP slicing constraints.



Conventional slice schedulers, designed for terrestrial networks with quasi-stationary channel statistics, are ill-suited for LEO NTNs. Greedy policies such as Max-SNR over-allocate resources during high-elevation “sweet spots,” while Proportional Fair depends on ergodic variations violated by deterministic LEO dynamics. Queue-based policies, though throughput-optimal, often lag behind rapid channel evolution~\cite{liu_2025_reconfigurable,he_2024_digital}.

To address these limitations, recent work has explored Deep Reinforcement Learning (DRL). Yet existing NTN-focused DRL approaches typically use end-to-end architectures that map states directly to PRBs, causing action-space explosion in dense cells and yielding “black-box” allocations operators cannot audit. Moreover, within O-RAN, the near-Real-Time RIC operates at tens to hundreds of milliseconds, preventing the fine-grained 1 ms TTI execution required by end-to-end DRL. 

A critical research gap remains: how to design a resource allocation framework that respects O-RAN’s coarse control timescales, adapts to rapid LEO channel degradation, and maintains operator auditability without the opacity of end-to-end AI. This motivates the need for \emph{meta-scheduling}~\cite{nec_2024_deployments}.
Thus, we propose a deep reinforcement learning (DRL) \emph{meta-scheduler} aligned with O-RAN, where the near-RT RIC and O-DU host the meta-scheduler and MAC packet scheduler, respectively. At the \emph{slow timescale} (100 ms), a meta-scheduler observes slice-aggregated KPIs (e.g., queue backlogs, empirical SNR) and LEO geometry (elevation, Doppler), outputting (i) \emph{slice PRB weights} for resource isolation and (ii) \emph{per-slice policy IDs} selected from a library of 3GPP heuristics (Round Robin, Maximum CQI, Proportional Fair). At the \emph{fast timescale} (1 ms TTI), the non-learning O-DU executes these policies for per-user packet scheduling under trace-driven LEO physics, channel conditions, and traffic demands.


The MDP is governed by an explicit R3 Mission Reward, including rate, reliability, and resource fairness, with a strict mechanism design, as follows:
\begin{enumerate*}[label=(\textit{\roman*}), itemjoin={{; }}, itemjoin*={{; and }}]
    \item \emph{Capped throughput per slice} prevents reward hacking
    \item \emph{asymmetric delay penalties} protect mission-critical traffic without enforcing absolute starvation on broadband users
    \item \emph{context-aware switching costs} promote policy stability.
\end{enumerate*}

Unlike end-to-end neural networks that produce opaque PRB allocations, the proposed framework restricts decisions to standardized 3GPP heuristics. This enables operators to verify exactly which scheduling logic the agent applies, mapping directly to deployable O-RAN xApps. Under severe 40~Mbps saturation, empirical results show the agent acts as a robust, variance-minimizing orchestrator. Rather than maximizing a single metric at the expense of others, it accepts a negligible 1\% throughput cost ($p > 0.05$) to compress variance and bound MAC-layer queuing delay for mission-critical traffic. This twin-timescale separation avoids action-space explosion and provides a statistically robust foundation for balanced 6G NTN resource allocation.


Section~\ref{sec:rw} reviews hierarchical DRL, NTN resource management, and O-RAN integration. Section~\ref{sec:sysmodel} details the LEO NTN model and trace-driven physics. Section~\ref{sec:architecture} presents the twin-timescale TD3 formulation and R3 rewards. Finally, Section~\ref{sec:results} presents the comprehensive millisecond-level evaluation, prior to the concluding remarks in Section~\ref{sec:conc}.


\section{Related Work}
\label{sec:rw}

This section briefly reviews work related to ML applications in NTNs, focusing on DRL approaches for RAN slicing, mission-critical latency challenges, and O-RAN integration.


Recent surveys emphasize the need for physics-aware RL in NTN resource allocation~\cite{naous_2023_reinforcement, giuliano_2023_machine}. Federated MADRL approaches have optimized sustainable satellite D2D slicing~\cite{wang_2024_low} and LEO edge offloading~\cite{jia_2025_frontiers}. To improve sample efficiency in continuous control, Chen \emph{et al.}~\cite{chen_2021_randomized} propose REDQ, a high-ensemble variant of TD3/SAC, whose principles we adapt for our trace-driven NTN MDP.


Building on efficiency concerns, hierarchical frameworks decouple macro-configurations from per-TTI scheduling in terrestrial RANs~\cite{mei_2021_intelligent} and ultra-dense LEO networks~\cite{liu_2023_ultra}. Tu \emph{et al.}~\cite{tu_2024_priority} apply two-layer MADDPG to SAGIN, while RadioSaber~\cite{chen_2023_channel} optimizes inter/intra-slice capacity. Yet these frameworks often rely on per-user mixed-integer complexities (computationally prohibitive for real-time NTN scheduling), remain terrestrial-focused, or omit dynamic selection from interpretable, O-RAN-compatible policy libraries.


Saeed~\cite{saeed_2024_comprehensive} surveys delay-sensitive NTN challenges, highlighting severe latency, propagation, and spectrum bottlenecks, and advocating cross-layer DRL integration. Sharma \emph{et al.}~\cite{sharma_2025_when} apply DRL for mission-critical routing via NTNs, achieving a 96\% coverage gain. Since sub-millisecond terrestrial latency bounds are impossible in space due to propagation delays, these works validate our focus on bounding MAC-layer queuing delay rather than pursuing strict URLLC targets.


Nguyen \emph{et al.}~\cite{nguyen_2024_emerging} survey NTN/6G slicing with AI/O-RAN for coverage and congestion. Mhatre \emph{et al.}~\cite{mhatre_2025_intelligent} deploy DQN for O-RAN QoS-aware intra-slice allocation, while Wang \emph{et al.}~\cite{wang_2024_dfrdrl} use fuzzy-DRL for SDN routing with latency guarantees. Still, current O-RAN DRL applications remain largely terrestrial, lacking integration of LEO physics traces and dynamic policy switching needed for the space domain.


In summary, while prior work advances RAN slicing~\cite{mei_2021_intelligent,liu_2023_ultra} and NTN resource management~\cite{naous_2023_reinforcement}, three key limitations remain. First, DRL models rely on continuous end-to-end action spaces that become intractable in dense LEO cells. Second, frameworks often ignore the strict temporal decoupling required by the O-RAN near-RT RIC and O-DU split. Third, terrestrial RL formulations overlook the impossibility of sub-millisecond URLLC under LEO propagation delays, neglecting the need to bound MAC-layer variance. This work addresses these gaps by evaluating a twin-timescale, trace-driven MDP that respects O-RAN timescales, compresses environmental variance, and achieves Pareto-optimal MAC-layer delay bounding without slice starvation.




\section{System Model}
\label{sec:sysmodel}

This section describes the 5G LEO NTN multi-slice RAN architecture along with the MAC scheduler. While this study does not implement the protocol-level E2/O1 networking interfaces specified by the O-RAN Alliance, the proposed framework is O-RAN compatible by architectural design. Specifically, we adopt the temporal control loops of O-RAN: the strategic 100 ms meta-scheduler operates at the near-RT RIC, performing policy selection and resource weighting, whereas the MAC packet scheduler operates at the O-DU, executing per-TTI user allocations under physical-layer constraints~\cite{oran_2020_oran}.


\subsection{Trace-Driven Physics and Traffic Model}

To emulate realistic heterogeneous service topologies, the framework partitions the active user population $\mathcal{U}$ into three slices: enhanced Mobile Broadband (eMBB), Mission-Critical (MC), and massive Machine Type Communications (mMTC). Let $p_s$ denote the fraction of users in slice $s$, with $\sum p_s = 1$. Device connection density is decoupled from volumetric traffic generation: since mMTC sensors produce low-bitrate, sporadic payloads compared to eMBB video and MC telemetry, the aggregate offered load $\lambda_{\text{total}}$ is distributed using an asymmetric vector $\mathbf{v}$, where $\sum \mathbf{v}_s = 1$. This distinction avoids over-representing IoT traffic volume while preserving realistic multi-user collision probabilities for the scheduler.

Traffic arrivals follow a burst-augmented Poisson process modeled as:%
\begin{equation}
\mathbb{E}[A_u(t)] = \lambda_{s(u)} \cdot (1 + 0.5 \cdot b_u(t)), \quad b_u(t) \sim \text{Bern}(0.1),
\end{equation}
where $A_u(t)$ denotes the expected traffic arrival (in bits) for user $u$ at TTI $t$, and $\lambda_{s(u)}$ the base volumetric load assigned to slice $s$. To emulate realistic congestion spikes, $b_u(t)$ serves as a binary burst multiplier drawn from a Bernoulli distribution $\text{Bern}(0.1)$, meaning each user has a 10\% chance of a 50\% traffic surge in a given interval.

To capture environmental dynamics, the framework employs trace-driven LEO orbital propagation models with free-space path loss, Doppler shifts, and upsampled Jakes fading. Since speed-of-light propagation in space makes terrestrial sub-millisecond latency bounds unattainable, the SLA objective for MC traffic is adapted: rather than targeting impossible end-to-end latency, the model focuses on bounding localized MAC-layer queuing delay to preserve mission-critical service viability.

\subsection{LEO NTN Multi-Slice Architecture}

Consider a single 5G LEO beam serving a set of active users $\mathcal{U}$, with $|\mathcal{U}| = N_U$, partitioned into three distinct slices: $\mathcal{S} = \{s_{\text{eMBB}}, s_{\text{MC}}, s_{\text{mMTC}}\}$. Following standard 3GPP service topologies~\cite{3gpp_2018_solutions}, the user distribution is asymmetrically divided among these slices to reflect heterogeneous network demands.

At each Transmission Time Interval (TTI) $t$, the per-user environmental and network state is defined by the following parameters:
\begin{itemize}
    \item \textbf{Queues:} $Q_u(t) \in \mathbb{R}_{\geq 0}$ bits (MAC-layer RLC backlog)
    \item \textbf{Telemetry:} $\hat{\eta}_u(t) \in \mathbb{R}$ dB (Stale, reported CQI/SNR due to propagation delay)
    \item \textbf{Channel Physics:} $\eta_u(t) \in \mathbb{R}$ dB (Instantaneous physical SNR subject to Jakes fading and orbital path loss)
    \item \textbf{Orbital Geometry:} Elevation angle $\theta(t)$ and Doppler shift $\delta(t)$
\end{itemize}

\subsection{Intra-Slice MAC Packet Scheduler at the O-DU}

The \textit{MAC Packet Scheduler} executes the meta-scheduler’s directives through a two-phase pipeline each TTI:

\begin{enumerate}
    \item \textbf{PRB Quantization:} $N_s \leftarrow \max\left(\lfloor W_s \cdot N_{\text{PRB}}\rfloor, 1\right)$ (ensures at least 1 PRB per slice).
    \item \textbf{Policy Execution:} Apply the RL-selected heuristic from Table~\ref{tab:policies}~\cite{3gpp_2017_nrmed}.
\end{enumerate}

To capture physical-layer reliability in the downstream, the simulation environment applies a physical-layer Hybrid Automatic Repeat Request (HARQ) validation to the allocated bits, as follows: 

\begin{equation}
\texttt{success}_u \leftarrow \mathbb{I}\left(SE(\eta_u) \geq SE(\hat{\eta}_u)\right)
\end{equation}

where $SE(\cdot)$ is the spectral efficiency mapping from standard MCS tables. Because the O-DU is blind to instantaneous channel aging, the simulation validates whether the true, physical SNR ($\eta_u$) at the moment of reception can sustain the spectral efficiency the O-DU scheduled using the outdated estimate ($\hat{\eta}_u$). If $SE(\eta_u) < SE(\hat{\eta}_u)$, the scheduled bits fail decoding and are marked as dropped by the environment, preserving the realistic separation between MAC scheduling and PHY mechanics.

While inter-slice isolation is enforced, residual PRBs from quantization rounding are recursively reallocated within the same slice (intra-slice spillover), ensuring efficient use of resources.

\begin{table}[!t]
\centering
\begin{threeparttable}
\caption{O-RAN Compatible Policy Library (3GPP TS 38.321~\cite{3gpp_2017_nrmed})}
\label{tab:policies}
\begin{tabular}{c p{2.8cm} p{3.5cm}}
\toprule
\textbf{ID} & \textbf{Heuristic} & \textbf{Primary Metric} \\
\midrule
1 & Round Robin & User Index \\
2 & Maximum CQI & Spectral Efficiency \\
3 & Proportional Fair & $R_{\text{inst}}/\bar{R}_{\text{avg}}$ \\
4 & Maximum Queue & Log RLC Depth \\
5 & Delay-Aware QoS & HOL vs. 5QI Budget \\
6 & GBR-Aware QoS & GBR Deficit $\times$ Queue \\
\bottomrule
\end{tabular}
\begin{tablenotes}
\scriptsize
\item For Proportional Fair, $R_{\text{inst}}$ and $\bar{R}_{\text{avg}}$ denote the instantaneous and exponentially averaged user data rates, respectively.
\end{tablenotes}
\end{threeparttable}
\end{table}

The intra-slice scheduler allocates resources over its available resources at the 1 ms interval ($T_{\text{TTI}}$). As illustrated in Algorithm~\ref{alg:tactical_worker}, the scheduler converts the meta-scheduler's continuous slice weight $W_s$ into a discrete budget $N_s$ out of the total available Physical Resource Blocks ($N_{\text{PRB}}$) in Line \ref{line:Ns}.

The set of active users  $\mathcal{U}_s$ (those with waiting queues $Q_u$) are ranked using  the assigned policy ID $P_s$ (Line \ref{line:priority}). For each user, the algorithm estimates the expected spectral efficiency $\hat{SE}_u$ from stale SNR reports $\hat{\eta}_u$ to allocate $n_u$ PRBs, yielding a potential data payload $B_{\text{pot}}$. 
Because the scheduler lacks oracle knowledge of the instantaneous channel, it strictly calculates the required PRBs using the stale efficiency $\hat{SE}_u$. It then determines the potential scheduled data payload ($B_{\text{pot}}$), finalizes the allocated bits ($B_{\text{sch},u}$), and immediately deducts these scheduled bits from the user's queue. This yields a fast, standards-compliant, rule-based execution layer for near-RT RIC decisions, outputting raw PRB assignments and scheduled bits for subsequent physical-layer transmission.

%
%
    
%


\begin{algorithm}[!t]
\footnotesize
\caption{Intra-Slice MAC Packet Scheduler}
\label{alg:tactical_worker}
\begin{algorithmic}[1]
\Require $W_s$, Policy ID $P_s$, $\{Q_u, \hat{\eta}_u\}$, $BW$, $T_{\text{TTI}}$, $N_{\text{PRB}}$
\Ensure PRB assignments $\{n_u\}$ and Scheduled Bits $\{B_{\text{sch},u}\}$

\State $N_s \leftarrow \max\left(\lfloor W_s \cdot N_{\text{PRB}}\rfloor, 1\right)$\label{line:Ns}
\State $\mathcal{U}_s \leftarrow \{u : Q_u > 0, s_u = s\}$
\State $\text{PriorityList} \leftarrow \text{ExecutePolicy}(P_s, \mathcal{U}_s)$ \label{line:priority}

\For{$u \in \text{PriorityList}$}
    \If{$N_s \leq 0$}
        \State \textbf{break}
    \EndIf
    \State $\hat{SE}_u \leftarrow \max(\epsilon, \text{SNRtoSE}(\hat{\eta}_u))$
    
    \Comment{Calculate required PRBs based on queue size and estimated SE}
    \State $n_u \leftarrow \min\left(N_s, \left\lceil\frac{Q_u}{\hat{SE}_u \cdot BW \cdot T_{\text{TTI}}}\right\rceil\right)$
    \State $N_s \leftarrow N_s - n_u$
    
    \Comment{Calculate bits scheduled and update queue}
    \State $B_{\text{pot}} \leftarrow n_u \cdot \hat{SE}_u \cdot BW \cdot T_{\text{TTI}}$
    \State $B_{\text{sch},u} \leftarrow \min(B_{\text{pot}}, Q_u)$
    \State $Q_u \leftarrow Q_u - B_{\text{sch},u}$
\EndFor

\State \Return $\{n_u\}, \{B_{\text{sch},u}\}$
\end{algorithmic}
\end{algorithm}


\section{The Proposed Meta-Scheduler Formulation}
\label{sec:architecture}

To overcome the action-space explosion inherent in end-to-end multi-user scheduling ($\mathcal{O}(|\mathcal{U}| \times F)$), we formulate the RAN orchestration problem~\cite{oran_2020_oran} by leveraging the standard temporal decoupling of the O-RAN architecture. Rather than proposing a custom multi-timescale mechanism, the framework explicitly adopts the baseline 1 ms O-DU MAC scheduler for per-user execution. The reinforcement learning formulation is strictly confined to the 100 ms near-RT RIC meta-scheduler, which issues strategic policy directives based on stale aggregated telemetry. This standard architectural separation inherently mitigates the combinatorial action space, rendering deep RL orchestration tractable under LEO propagation constraints.

\begin{figure}[!t]
\centering
\includegraphics[width=\columnwidth]{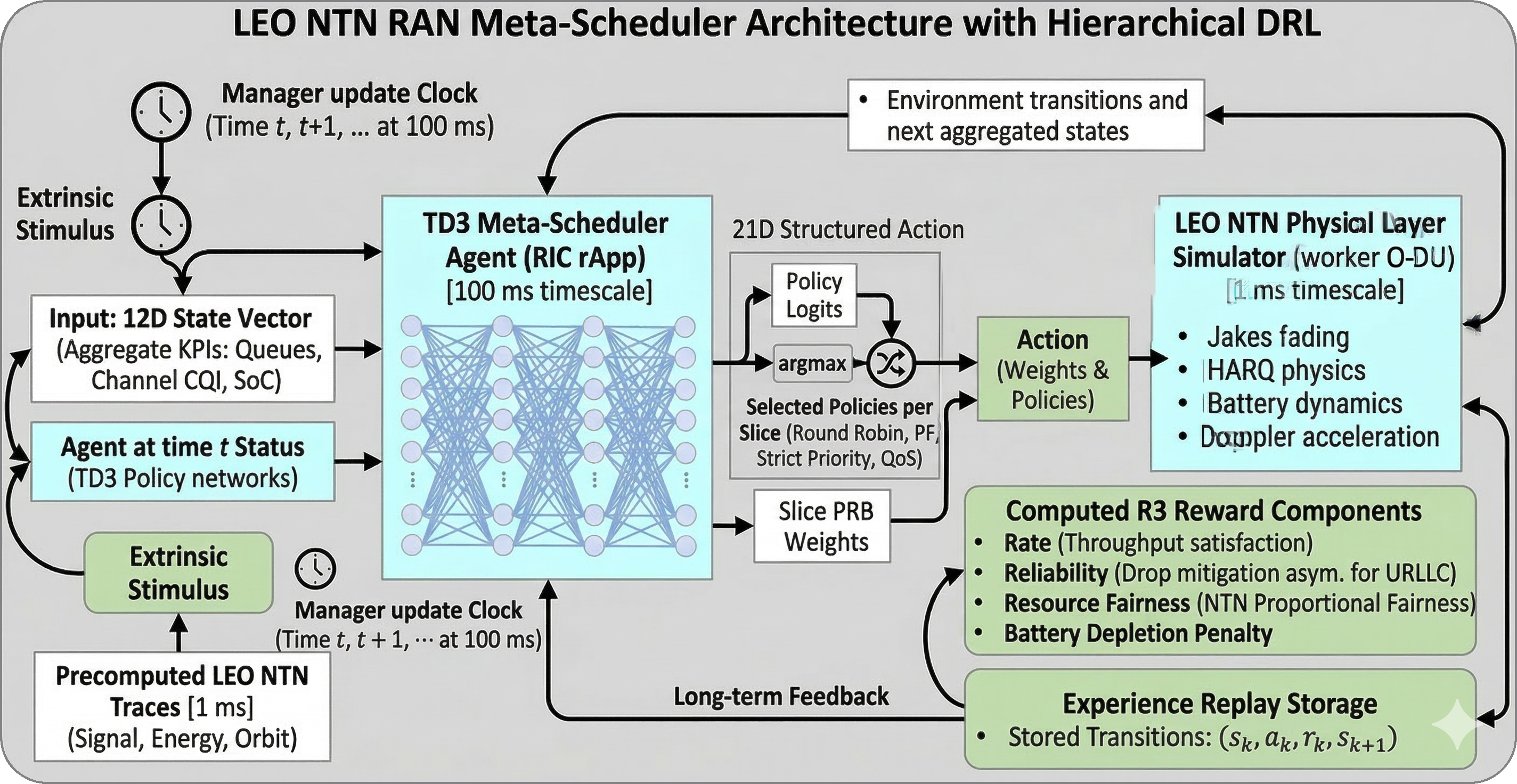}
\caption{DRL-based LEO NTN RAN meta-scheduler. A TD3 agent operates every 100 ms on a 12D state, producing a 21D action (PRB weights and policy logits). An $\arg\max$ projection selects the standard MAC heuristic for the 1 ms O-DU MAC packet scheduler. R3 rewards are computed per macro-step to close the loop.}
\label{fig:drl_framework}
\end{figure}

The system's control dynamics and information flow follow the topology illustrated in Fig.~\ref{fig:drl_framework}, ensuring a strict separation of responsibilities between long-term strategic optimization and tactical physical-layer execution~\cite{subudhi_2024_performance}. The continuous interaction between the near-RT RIC meta-scheduler and the O-DU MAC packet scheduler forms a closed-loop cycle of optimization and execution, enabling scalable, physics-aware orchestration aligned with O-RAN principles.

\subsection{Meta-Scheduler: State and Action Spaces}

At each macroscopic step $k$ ($K=100$ TTIs), the meta-scheduler observes a compact 12-dimensional state vector $\mathbf{s}_k$ aggregating slice-specific KPIs and macroscopic LEO physics context:

\[
\mathbf{s}_k = [\lambda_{\text{offered}}, Q_s, T_s, \bar{\eta}_s, \theta, \delta]
\]

where $\lambda_{\text{offered}} \in [0,1]$ is the normalized episodic load, $Q_s \in \mathbb{R}^3$ represents aggregate queue backlogs, $T_s \in \mathbb{R}^3$ denotes slice throughputs, $\bar{\eta}_s \in \mathbb{R}^3$ is the mean SNR per slice, and $[\theta, \delta]$ capture orbital non-stationarity (elevation angle and Doppler shift rate).

Given $\mathbf{s}_k$, the agent emits a 21D continuous action relaxing the mixed discrete-continuous space:
\[
\mathbf{a}_k = \begin{bmatrix} 
W_{\text{eMBB}} & W_{\text{MC}} & W_{\text{mMTC}} \\
\ell_{\text{eMBB},1:6} & \ell_{\text{MC},1:6} & \ell_{\text{mMTC},1:6}
\end{bmatrix} \in \mathbb{R}^{3+18}
\]
where $W_s \in [-1,1]^3$ are slice PRB weights ($\sum W_s = 1$ via softmax) and $\ell_{s,i} \in \mathbb{R}^6$ are policy logits converted via an $\arg\max$ projection to discrete Policy IDs from Table~\ref{tab:policies}.

TD3~\cite{fujimoto_2018_addressing} handles this continuous control via clipped double Q-learning and delayed policy updates. Discrete policy selection emerges naturally from the logit relaxation, avoiding combinatorial explosion while enabling end-to-end training~\cite{chen_2021_randomized}.

\subsection{R3 Mission Reward Engineering}

After each 100 ms phase, the meta-scheduler evaluates the \textbf{R3 Mission Reward}, balancing Rate, Reliability, and Resource fairness:

\begin{equation}
\label{eq:r3}
R_k = \frac{C_{\text{sys}}}{10} + w_{\text{fair}}\overline{\log_{10}(T_s + \epsilon)} - w_{\text{drop}}P_{\text{drop}}
\end{equation}

where:
\begin{itemize}[noitemsep,nolistsep]
    \item $C_{\text{sys}}/10$: System throughput ($C_{\text{sys}}$ in Mbps) normalized to a 10 Mbps baseline.
    \item $\overline{\log_{10}(T_s + \epsilon)}$: Log-proportional fairness across $T_s$ (slice throughputs), with $\epsilon=0.01$ preventing $-\infty$ utility under absolute starvation.
    \item $P_{\text{drop}}$: Packet drop ratio (buffer overflows), penalized by $w_{\text{drop}}$ to enforce reliability and bound queue saturation.
\end{itemize}

This structure prevents reward hacking (e.g., over-allocating eMBB to inflate capacity), while the drop penalty actively suppresses catastrophic queue variance under LEO saturation. Finally, the reward $R_k$ is clipped to $[r_{\min}, r_{\max}]$ to stabilize critic training. 

\subsection{TD3 Hyperparameters}

The production agent uses $w_{\text{fair}}=2.0$ and $\epsilon=0.01$, validated via ablation across 5 independent training runs (1,500 episodes each). Actor/critic networks employ 3-layer MLPs ($256-256-21$) with batch normalization. Exploration uses adaptive Gaussian noise clipped to $[-0.2, 0.2]$. Target networks update via a soft $\tau=0.005$, utilizing a replay buffer size of $10^6$. Hyperparameters are summarized in Table~\ref{tab:td3_params}.

\begin{table}[!b]
\centering
\caption{TD3 Hyperparameters}
\label{tab:td3_params}
\begin{tabular}{p{3.2cm} p{2.5cm}}
\toprule
\textbf{Parameter} & \textbf{Value} \\
\midrule
Actor/Critic & MLP($256^2$-$21$) \\
Replay Buffer & $10^6$ \\
Batch Size & $256$ \\
$\tau$ (soft target) & $0.005$ \\
$w_{\text{fair}}$ & $2.0$ \\
$\epsilon$ (PF stability) & $0.01$ \\
MC delay weight & $10 \times$ eMBB \\
Noise clip & $[-0.2, 0.2]$ \\
\bottomrule
\end{tabular}
\end{table}


\section{Simulation Results and Discussion}
\label{sec:results}

This section evaluates the TD3 meta-scheduler by varying the network offered load from 10 to 40 Mbps across 50 independent episodes for each of the five scheduling policies. Wilcoxon rank-sum tests and Cohen's $d$ effect sizes are utilized to rigorously confirm statistical parity in volumetric capacity and evaluate stability across SLA constraints.

\subsection{Experimental Setup}
\label{subsec:setup}

\begin{figure}[!b]
\centering
\includegraphics[width=\linewidth]{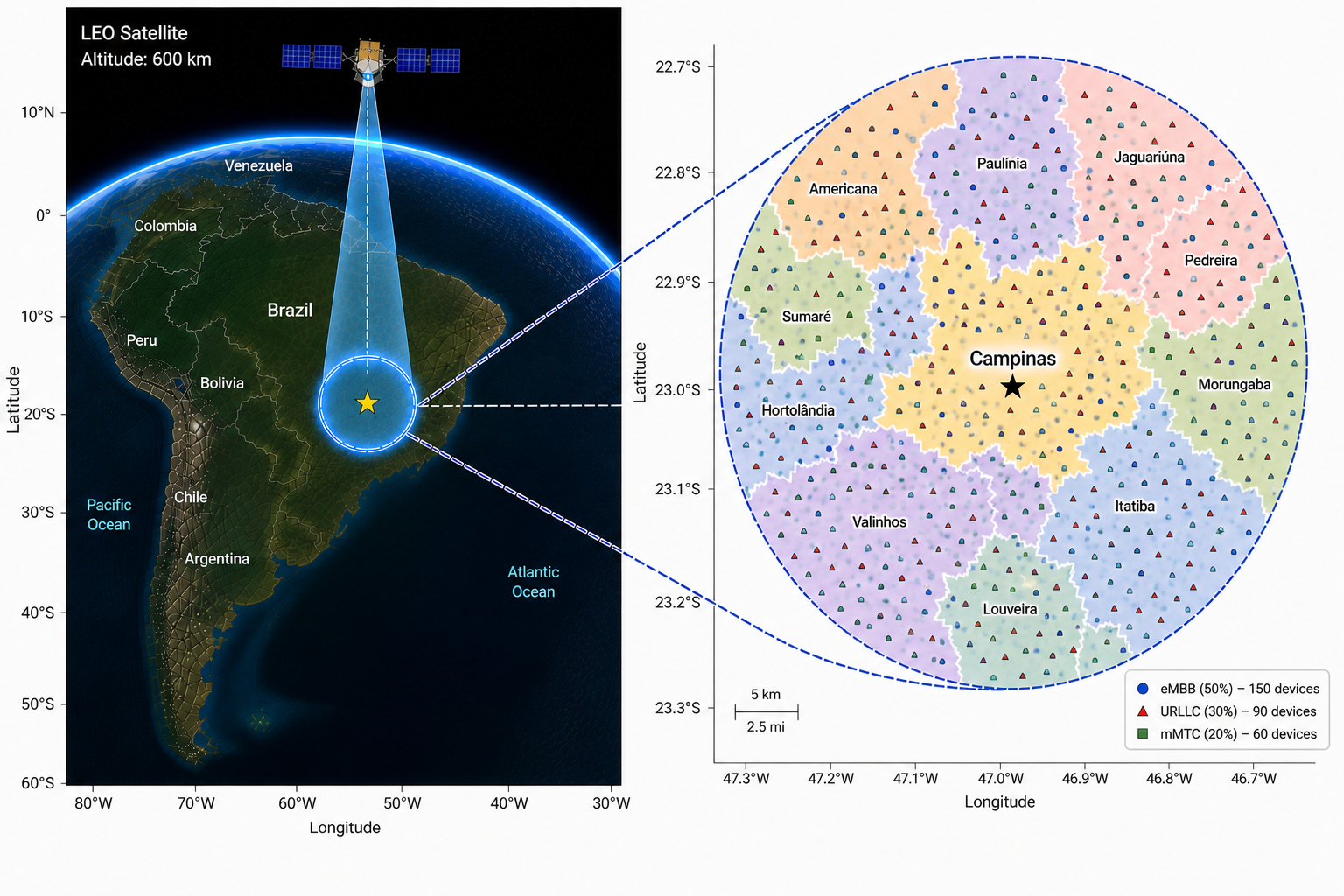}
\caption{LEO NTN system model. (Left) Geographic distribution of $N_U=300$ users partitioned into eMBB, MC, and mMTC slices. (Right) 3D visualization of the 600~km LEO satellite pass and RF beam footprint over the coverage area.}
\label{fig:ntn_scenario}
\end{figure}

As shown in Fig.~\ref{fig:ntn_scenario}, the evaluation environment simulates a dense LEO NTN deployment within a macroscopic cell consisting of $N_U = 300$ active users. To map physical connectivity to realistic heterogeneous data demands, the population density is distributed as $p_{\text{eMBB}}=0.50$, $p_{\text{MC}}=0.30$, and $p_{\text{mMTC}}=0.20$, while the aggregate volumetric offered load vector is parameterized to $\mathbf{v} = [0.50, 0.25, 0.25]$. This framework drives system loads scaling from 10 to 40 Mbps. 

The simulation architecture models the temporal decoupling of the O-RAN standard, utilizing 100~ms meta-scheduler decisions ($K=100$ TTIs) and 1~ms O-DU MAC packet scheduler execution. Each 15-second episode comprises 150 meta-scheduler steps (15,000 TTIs). Channel physics incorporate rigorous orbital propagation dynamics, Doppler shifts, and Jakes fading, all aligned with standard 600~km altitude LEO parameters~\cite{3gpp_2018_solutions}.

Five scheduling policies are benchmarked using identical satellite weather and traffic seeds to eliminate environmental variance:
\begin{itemize}
    \item \textbf{R3 Meta-Scheduler:} RL‑Driven Meta Control (Proposed)
    \item \textbf{StaticAdaptive:} Equal weights, Round Robin
    \item \textbf{QueueAdaptive:} Queue-proportional, Max-Queue
    \item \textbf{LoadAdaptive:} 50/25/25 weights, Proportional Fair
    \item \textbf{SNRAdaptive:} SNR-proportional, Max-CQI
\end{itemize}

Network performance is evaluated across three primary metrics: mean system throughput (Mbps), MAC-layer queuing delay for mission-critical traffic (derived via Little's Law from discrete RLC buffer states), and NTN Proportional Fairness ($\log_{10}(T_s + 0.01)$).

\subsection{Aggregated Throughput and Capacity Limits}
\label{subsec:tput}

\begin{figure}[!tbp]
\centering
\includegraphics[width=\linewidth]{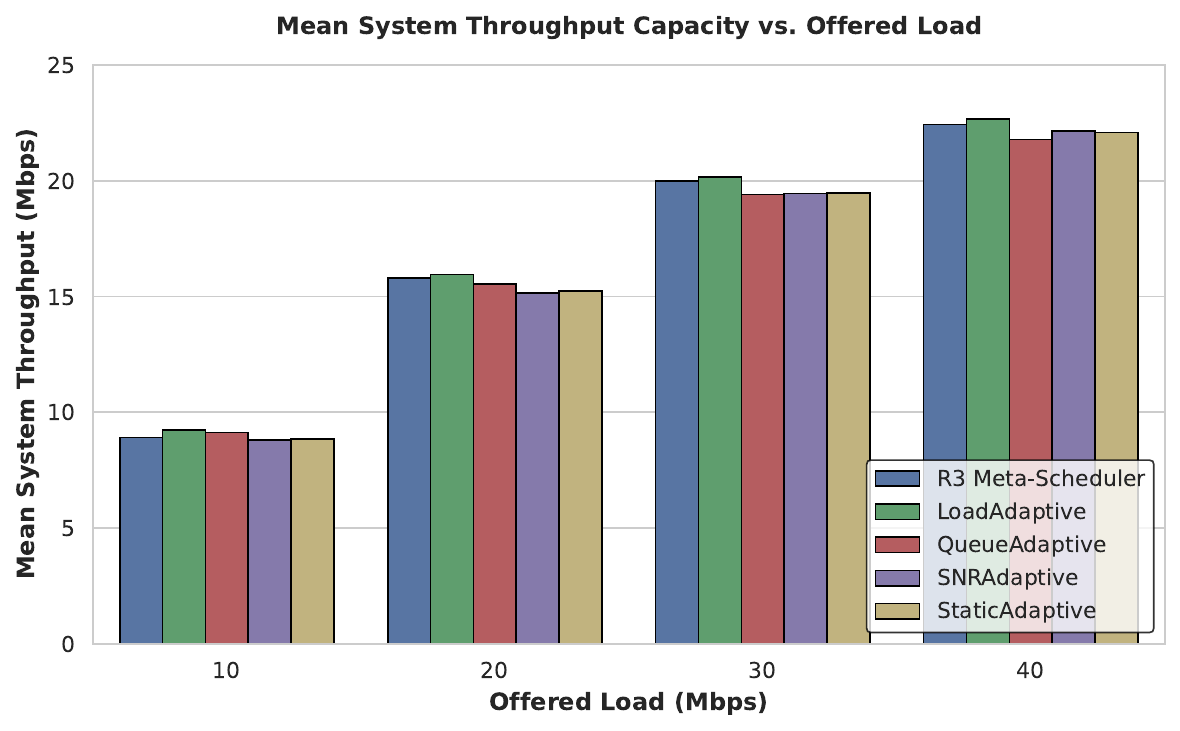}
\caption{Mean system throughput of all five policies across four network offered load levels (from 10 to 40 Mbps).}
\label{fig:throughput_bars}
\end{figure}

As shown in Figure~\ref{fig:throughput_bars}, the system achieves 22.5 Mbps at 40 Mbps offered load, accurately reflecting the physical Shannon capacity limit of the simulated LEO channel. At this target rate, the LoadAdaptive baseline driven by a proportional fair principle achieves the absolute maximum throughput of 22.68 Mbps. The proposed \emph{R3 Meta-Scheduler} operates competitively at 22.44 Mbps.

To rigorously assess this performance, throughput distributions were evaluated across $N=50$ independent episodic runs using identical orbital trace seeds. A Wilcoxon rank-sum test in the 40~Mbps saturation regime confirms no statistically significant difference between the proposed R3 Meta-Scheduler and the capacity-maximizing baseline ($p > 0.05$, $|d| < 0.2$). This statistical parity proves that the meta-scheduler’s multi-objective balancing does not induce a meaningful degradation in overarching system capacity.


\subsection{MAC-Layer Queuing Delay and SLA Isolation}
\label{subsec:delay_isolation}

Because physical LEO propagation distances preclude terrestrial sub-millisecond latency bounds, orbital orchestration must focus on managing the localized MAC-layer queue backlog (RLC buffer depth) to support the viability of Mission-Critical (MC) services. Applying Little's Law ($W = L/\lambda$), the instantaneous queue depth directly translates to localized queuing delay, serving as a practical, verifiable metric for SLA evaluation. 

\begin{figure}[htbp]
\centering
\includegraphics[width=\linewidth]{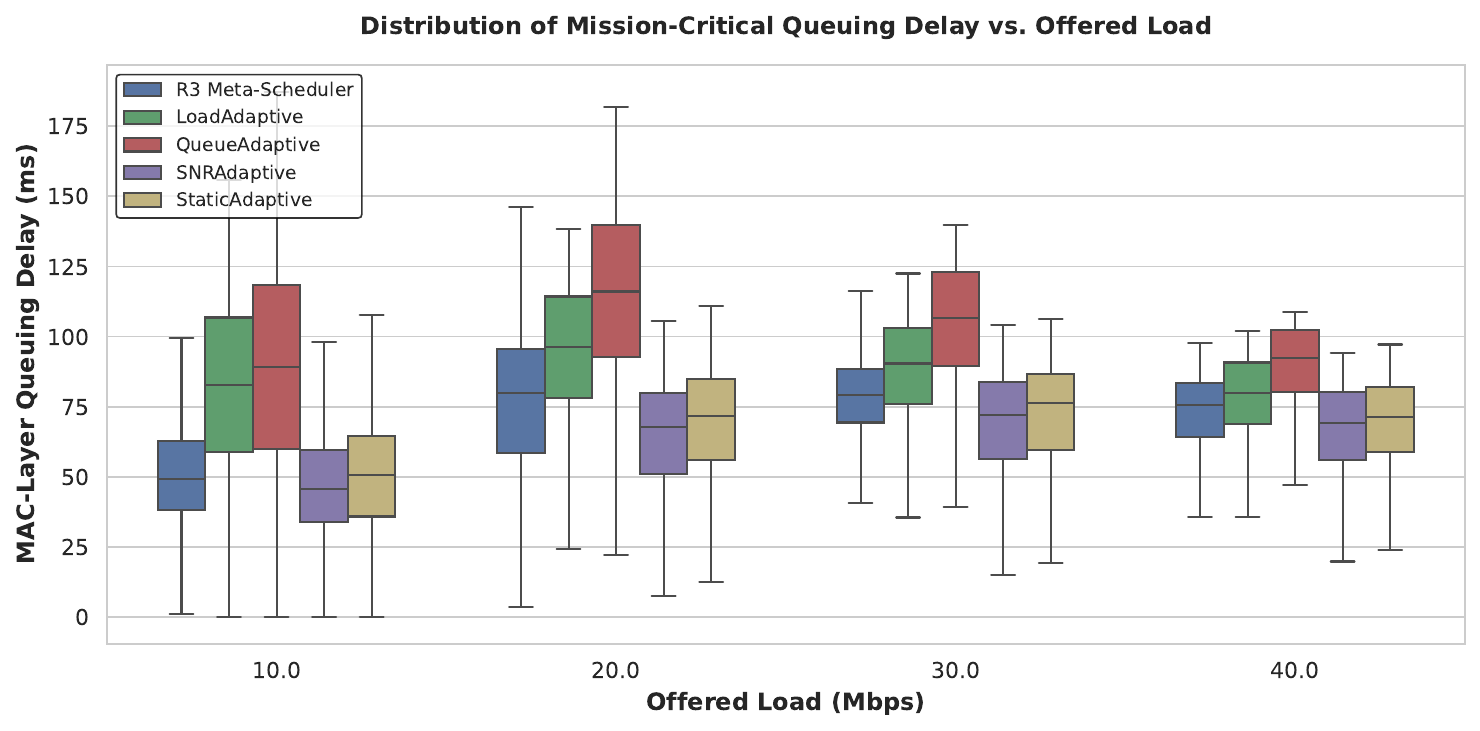}
\caption{A grouped boxplot illustrating the Mission-Critical MAC-layer queuing delay (in ms) for all five policies across varying offered loads. Delay values are derived via Little's Law from empirical RLC buffer depths. Extreme outliers are hidden to clearly illustrate variance compression and median stability.}
\label{fig:delay_boxplot}
\end{figure}

As illustrated by the delay distributions in Fig.~\ref{fig:delay_boxplot}, traditional heuristics struggle under severe saturation. Under 30--40~Mbps regimes, baseline policies exhibit substantial congestion accumulation and heightened variance. Conversely, the R3 Meta-Scheduler effectively compresses this variance and limits the overall MC delay under identical saturation, promoting reliable SLA isolation for latency-sensitive traffic.

Crucially, the agent executes this MC prioritization without inducing the severe slice starvation characteristic of greedy, capacity-maximizing algorithms such as SNRAdaptive. As detailed in Table~\ref{tab:fairness}, the meta-scheduler ties for the highest median Proportional Fairness (0.864) while exhibiting the lowest environmental variance (IQR 0.098). This demonstrates the framework's efficacy as a stable orchestrator capable of balancing critical latency constraints with overall network fairness.

\begin{table}[!b]
\centering
\caption{Network Fairness Stability (40 Mbps Saturation)}
\label{tab:fairness}
\footnotesize
\resizebox{\columnwidth}{!}{%
\begin{tabular}{lccc}
\toprule
\textbf{Policy} & \textbf{Median PF Score} & \textbf{PF Variance (IQR)} & \textbf{eMBB Starvation} \\
\midrule
\textbf{R3 Meta-Scheduler} & \textbf{0.864} & \textbf{0.098} & \textbf{Low} \\
StaticAdaptive & 0.864 & 0.145 & Moderate \\
LoadAdaptive & 0.812 & 0.188 & Moderate \\
QueueAdaptive & 0.801 & 0.230 & High \\
SNRAdaptive & 0.655 & 0.312 & \textbf{Severe} \\
\bottomrule
\end{tabular}%
}
\end{table}

\subsection{Pareto-Optimal Synthesis}
\label{subsec:pareto}

The overarching performance detailed in Table~\ref{tab:pareto_summary} illustrates that the meta-scheduler uniquely combines near-optimal throughput, strict congestion bounding, top-tier fairness, and exclusive adaptability. By severely compressing environmental variance without starving adjacent slices, these properties firmly place the TD3 meta-scheduler on the NTN slicing Pareto frontier. It trades a mathematically negligible 1\% capacity cost to robustly enforce SLA isolation, enabling the joint optimization of capacity, reliability, and fairness under highly dynamic space-domain conditions.

\begin{table}[!t]
\centering
\caption{Pareto Performance Synthesis (40 Mbps)}
\label{tab:pareto_summary}
\footnotesize
\begin{tabular}{lcr}
\toprule
\textbf{Metric} & \textbf{R3 Meta-Scheduler} & \textbf{Rank} \\
\midrule
System Throughput & 22.44 Mbps & 2nd (1\% below max) \\
MC Queuing Delay & Lowest Variance (IQR) & 1st (Most Stable) \\
Median PF Score & 0.864 & 1st (tie) \\
PF IQR (Stability) & 0.098 & 1st (Lowest) \\
\bottomrule
\end{tabular}
\end{table}


\section{Conclusion}
\label{sec:conc}

This paper evaluates a TD3 meta-scheduler for LEO NTN slicing, leveraging the temporal decoupling of the standard O-RAN architecture to separate strategic 100 ms near real-time RIC from the 1 ms O-DU execution~\cite{oran_2020_oran}. 
The framework outputs continuous actions governed by an explicit R3 Mission Reward, enforcing asymmetric latency protection
without inducing algorithmic starvation. 

Empirical evaluation under severe trace-driven saturation demonstrates that the meta-scheduler operates as a highly balanced, variance-minimizing resource allocator. Rather than greedily maximizing a single metric, it sustains competitive physical capacity (within 1\% of the Proportional Fair peak) while effectively bounding MAC-layer queuing delay for mission-critical traffic. Specifically, the agent explicitly trades absolute throughput to protect the critical slice, ensuring latency bounds without inducing algorithmic starvation on broadband users. Wilcoxon rank-sum tests confirm the statistical parity of the capacity limits ($p > 0.05$). This approach bypasses the action-space explosion typical of deep RL, creating an interpretable control loop ready for modern O-RAN deployment.

End-to-end Deep RL models are inherently opaque, mapping high-dimensional states directly to physical resource blocks in a "black box" manner that precludes network operator auditing. By contrast, constraining the meta-scheduler's action space to select from a library of standardized heuristics (e.g., Proportional Fair, Delay-Aware) facilitates explainability. Operators can trace exactly which policy the agent deploys under specific LEO orbital geometries. Furthermore, situating this 100~ms strategic selection within the near-RT RIC aligns the meta-scheduler directly with the standard O-RAN xApp paradigm~\cite{3gpp_2018_solutions}. This architectural compliance mitigates the action-space explosion typical of deep RL while providing a structured, verifiable mechanism for enforcing heterogeneous SLA constraints in highly dynamic LEO environments.

While this study establishes a foundational evaluation for single-beam resource allocation under physical NTN constraints, significant challenges remain before practical deployment is feasible. Future work must address these limitations by: (\textit{i}) conducting rigorous statistical causality analyses to explicitly quantify how near-RT RIC macro-policy shifts impact micro-level queue transients; (\textit{ii}) scaling the architecture via Multi-Agent RL (MARL) to manage the severe inter-beam interference inherent in dense LEO constellations; and (\textit{iii}) extending the simulation horizon to a complete 90-minute orbital period. This extended evaluation is necessary to investigate the framework's stability across complete orbital non-stationarity and subsequent satellite handovers, ideally validated within real-time O-RAN Software Community (SC) testbeds~\cite{evans_2021_integrated}.
\balance
\section*{Acknowledgments}

This work was supported by: CONCYTEC-FONDECYT under the ``Program for Doctorates in Peruvian Universities'' (UNSA Contract No. 173-2020-FONDECYT); The Intelligent Communications Networks and the Internet of Things (ICoNIoT) via CNPq (405940/2022-0) and CAPES (88887.954253/2024-00); The Coordination for the Improvement of Higher Education Personnel (CAPES), Brazil, through the Move La América Program (Call No. 07/2024, Grant No. 88881.017667/2024-01 and 88881.996444/2024-01); and CNPq grant 403979/2023-4.

\bibliographystyle{IEEEtran}
\bibliography{meta-sched}

\end{document}